\documentclass[onecolumn,preprintnumbers,elsart]{revtex4}
\usepackage{makeidx}
\usepackage{amssymb}
\usepackage{amsmath}
\usepackage{mathrsfs}
\usepackage{graphicx}
\usepackage{dcolumn}
\usepackage{bm}
\usepackage[center]{subfigure}
\usepackage{color}

\begin{document}

\title{Immiscible and miscible states in binary condensates in the ring
geometry}
\author{Zhaopin Chen$^{1}$}
\author{Yongyao Li$^{1,2}$}
\author{Nikolaos P. Proukakis$^{3}$}
\author{Boris A. Malomed$^{1}$}
\affiliation{$^{1}$Department of Physical Electronics, School of Electrical Engineering,
Faculty of Engineering, Tel Aviv University, Tel Aviv 69978, Israel\\
$^{2}$School of Physics and Optoelectronic Engineering, Foshan University,
Foshan 52800, China\\
$^{3}$Joint Quantum Centre (JQC) Durham-Newcastle, School of Mathematicss,
Statistics and Physics, Newcastle University, Newcastle upon Tyne NE1 7RU,
England, United Kingdom}

\begin{abstract}
We report detailed investigation of the existence and stability of mixed and
demixed modes in binary atomic Bose-Einstein condensates with repulsive
interactions in a ring-trap geometry. The stability of such states is
examined through eigenvalue spectra for small perturbations, produced by the
Bogoliubov-de Gennes equations, and directly verified by simulations based
on the coupled Gross-Pitaevskii equations, varying inter- and intra-species
scattering lengths so as to probe the entire range of
miscibility-immiscibility transitions. In the limit of the one-dimensional
(1D) ring, i.e., a very narrow one, stability of mixed states is studied
analytically, including hidden-vorticity (HV) modes, i.e., those with
opposite vorticities of the two components and zero total angular momentum.
The consideration of demixed 1D states reveals, in addition to stable
composite single-peak structures, double- and triple-peak ones, above a
certain particle-number threshold. In the 2D annular geometry, stable
demixed states exist both in radial and azimuthal configurations. We find
that stable radially-demixed states can carry arbitrary vorticity and,
counter-intuitively, the increase of the vorticity enhances stability of
such states, while unstable ones evolve into randomly oscillating angular
demixed modes. The consideration of HV states in the 2D geometry expands the
stability range of radially-demixed states.
\end{abstract}

\maketitle

\section{Introduction}

Superfluid mixtures are currently routinely probed in experiments with
ultracold atomic gases. In addition to Bose-Bose mixtures of different
isotopes and atomic species \cite{myatt_burt_1997, hall_matthews_1998,
maddaloni_modugno_2000,mertes_merrill_2007,papp_pino_2008,
sugawa_yamazaki_2011, modugno_modugno_2002, thalhammer_barontini_2008,
mccarron_cho_2011, lercher_takekoshi_2011, pasquiou_bayerle_2013,
wacker_jorgensen_2015, Petrov_2015, wang_li_2016,
Cabrera_2018,Cabrerera2,Semeghini_2018,Semeghini2}, experimentalists have in
the past few years created condensates with a spin degree of freedom~\cite%
{stamper-kurn_ueda_2013}, also implementing spin-orbit coupling which gives
rise to exciting new states \cite{Galitski_2013, Zhou_2013,Goldman_2014,
Zhai_2015, Malomed_2018}; moreover, recent achievements have led to the
generation of doubly-superfluid Bose-Fermi mixtures~\cite%
{ferrier-barbut_delehaye_2014}, in which both components are condensed, a
state so far inaccessible in other settings (such as superfluid helium).
Although the stability and phase diagrams of such systems have been
extensively studied in the course of more than 20 years ~\cite%
{esry_greene_97,pu_bigelow_1998,ao_chui_98,timmermans_98,ohberg_1999,trippenbach_goral_2000,delannoy_murdoch_2001,riboli_modugno_02, jezek_capuzzi_02,svidzinsky_chui_2003,kasamatsu_tsubota_2004,ronen_bohn_2008,takeuchi_ishino_2010,suzuki_takeuchi_2010,mason_aftalion_2011,aftalion_mason_2012,wen_liu_12, pattinson_billam_2013,hofmann_natu_2014,pattinson_parker_2014,edmonds_lee_2015a,edmonds_lee_2015b,lee_proukakis_2016, liu_pattinson_2016,lee_jorgensen_2016,lee_jorgensen_2018}%
, even simple hetero-species binary mixtures still reveal unexpected
features, such as the role of the trap sag, atom number and kinetic energy
contribution to the extent of miscibility in trapped configurations \cite%
{lee_jorgensen_2016,wen_liu_12,Richaud_2019}, and nontrivial effects of the expansion on
the mixtures' dynamics \cite{lee_jorgensen_2018,Yongyao_2018}.
%[spin-drag; spin-superfluidity]

Configurations which keep drawing growing interest in studies of ultracold
atomic gases are based on the annular, alias ring-trap, geometry
%One possible way to realized this system is by using an strong laser beam with annular shape acting on the binary BECs. In experiment, creation of toroidal trap are already available, and there are various ways to construct
%\cite{Ryu,Arnold,Gupta,Amico,}.
\cite{gupta_murch_2005,arnold_garvie_2006, ryu_andersen_2007,
ramanathan_wright_2011, eckel_lee_2014, corman_chomaz_2014, kumar_eckel_2017}%
. These configurations are interesting as they lead to closed geometries
with controlled flows, that are also of potential use to the emerging field
of atomtronics \cite{eckel_lee_2014, Amico_2017}. In this context, mixtures
of atomic condensates in toroidal traps and the possibility of sustaining
stable persistent currents in them have been previously considered in \cite%
{smyrnakis_bargi_2009, shimodaira_kishimoto_2010, bargi_malet_2010,
anoshkin_wu_2013, yakimenko_isaieva_2013, mason_2013, abad_sartori_2014,
white_hennessy_2016, Yakimenko_2015, white_zhang_2017}, and the
corresponding experimental observation \cite{beattie_moulder_2013} has
helped to clarify some issues, also raising new questions, such as expansion
of the stability area for such states. The aim of the present work is to
perform a full classification of accessible stable mixture states in such a
geometry, including examination of their stability and decay channels of
their unstable counterparts, both in the absence and presence of overall
rotation. Given the potential significance of multi-component states in
ring-shaped traps for applications such as rotational sensors, such a
classification is relevant. It can also assist in developing methods for
control of such mixtures in the experimental work which is currently going
on in many laboratories.

Specifically, in this work, we construct a binary Bose-Einstein-condensate
(BEC) system trapped in an annular geometry, through the mean-field analysis
in the presence of inter- and intra-species interactions, whose parameters
are varied in broad limits. After analyzing the corresponding
one-dimensional (1D) problem, we focus on the more experimentally-relevant
2D annular structure. We implement periodic boundary conditions (b.c.) in
the azimuthal direction in the 1D case, and zero b.c. at inner and outer
boundaries of the 2D annular structure. The latter b.c. set enables one to
study how the annular structure affects density patterns in the repulsive
bosonic mixtures, as a result of the existence of different demixed and
mixed states, and their stability.

This paper is structured as follows. First, Sec. II introduces and
analytically considers our basic model for the mixtures in both 1D and 2D
geometries, and presents analytical results for spatially uniform 1D mixed
solutions, with zero and hidden vorticities (HV), the latter implying
opposite topological charges in the two components, which makes it possible
to construct stable binary vortex states with zero total angular momentum in
nonlinear optics \cite%
{desyatnikov_kivshar_2001,desyatnikov_2005,bigelow_park_2002,Leykam_2013,Salgueiro_2016}
and BEC \cite{mihalache_mazilu_2006, brtka_gammal_2010, Yakimenko_2012,
He_2012, Linghua_2013, Ishino_2013, Hoashi_2016, Yongyao_2018}. Most
essential are analytical results for stability of these states. Sec. III
presents the key results, showing various types of mixed and demixed states
in 1D, characterized by different numbers of peaks in them, and both mixed
and demixed 2D states. The latter ones include both radially-demixed modes,
with different vorticities, and their azimuthally-demixed counterparts. Such
states are obtained by means of the imaginary-time-propagation method,
applied to the coupled Gross-Pitaevskii equations (GPEs). We also address
effects of the strength of the repulsive intra-component interaction,
annular width, and embedded vorticity on the existence and stability of
different states. A noteworthy finding is that the stable radially-demixed
states can exist with arbitrary vorticity. Our findings are summarized in
Sec.~IV.

\section{The mean-field models}

At low temperatures, a binary condensate mixture is well described by the
mean-field theory for the set of wave functions $\phi $ and $\psi $ of the
two components. Here we address the system (e.g., a heteronuclear one) which
does not admit linear interconversion (Rabi and/or spin-orbit coupling)
between the components. The wave functions obey the system of GPEs with
nonlinear terms accounting for self- (intra-species) and cross-
(inter-species) interactions. In the normalized form, the GPE system is
written as
\begin{eqnarray}
i\phi _{t}+\frac{1}{2m_{1}}\nabla ^{2}\phi -\left( g_{1}\left\vert \phi
\right\vert ^{2}+g_{12}|\psi |^{2}\right) \phi &=&0,  \notag \\
&&  \label{phipsi} \\
i\psi _{t}+\frac{1}{2m_{2}}\nabla ^{2}\psi -\left( g_{2}\left\vert \psi
\right\vert ^{2}+g_{12}|\phi |^{2}\right) \psi &=&0,  \notag
\end{eqnarray}%
where $m_{1,2}$ are scaled atomic masses, $g_{1,2}$ are coefficients of
self-interaction in species $\phi $ and $\psi $, and $g_{12}>0$ is the
cross-interaction coefficient. In this work, the analysis is restricted to
repulsive interactions, with $g_{1,2,12}>0$. Then, condition $g_{12}=\sqrt{%
g_{1}g_{2}}$ separates the mixing ($\sqrt{g_{1}g_{2}}>g_{12}$) and
phase-separation (demixing,$\sqrt{g_{1}g_{2}}<g_{12}$) regimes in free space
\cite{Mineev_1974}. This criterion is modified by the presence of a trapping
potential, which tends to enhance the miscibility \cite{Merhasin_2005,
lee_jorgensen_2016, wen_liu_12}.

Equations (\ref{phipsi}) are supplemented by b.c. set at rigid edges, $r=r_{%
\mathrm{outer}}$ and $r=r_{\mathrm{inner}}$ of the annular area filled by
the condensate ($r$ is the radial coordinate):%
\begin{equation}
\phi \left( r=r_{\mathrm{outer,inner}}\right) =\psi \left( r=r_{\mathrm{%
outer,inner}}\right) =0.  \label{r}
\end{equation}%
By means of scaling, we fix
\begin{equation}
r_{\mathrm{inner}}\equiv 1,  \label{inner}
\end{equation}%
and define the annulus' width,%
\begin{equation}
w\equiv r_{\mathrm{outer}}-1.  \label{w}
\end{equation}%
These b.c. imply that the annular area is confined by rigid circular
potential walls, as in a recent experiment \cite{Hueck_2018} (performed for
a gas of fermions).

The total norm of the 2D system is%
\begin{equation}
N=\int \int \left( |\phi |^{2}+|\psi |^{2}\right) dxdy\equiv N_{\phi
}+N_{\psi },  \label{NN}
\end{equation}%
where the integration is performed over the annular region, or over the
circumference, in the 1D limit, which corresponds to very tight confinement
in the radial direction [see Eq. (\ref{N})] below. The energy (Hamiltonian)
of the coupled system is
\begin{equation}
E=\int \int \left[ \frac{1}{2m_{1}}|\nabla \phi |^{2}+\frac{1}{2m_{2}}%
|\nabla \psi |^{2}+\frac{1}{2}(g_{1}|\phi |^{4}+g_{2}|\psi
|^{4})+g_{12}|\phi |^{2}|\psi |^{2}\right] dxdy,  \label{1_hamilt}
\end{equation}%
which is accordingly reduced in the 1D limit.

\subsection{The one-dimensional setting}

To define the 1D limit, we assume that the single coordinate, $x$, running
along the ring of radius $r=1$ [which is fixed by scaling in agreement with
Eq. (\ref{inner})], takes values $0\leq x\leq 2\pi $. Then, the substitution
of solutions in the Madelung form,
\begin{equation}
\phi \left( x,t\right) =a\left( x,t\right) \exp \left( i\chi \left(
x,t\right) \right) \,\,\,,\,\,\,\psi \left( x,t\right) =b\left( x,t\right)
\exp \left( i\eta \left( x,t\right) \right) ,  \label{Madelung}
\end{equation}%
leads to the system of four real equations for the amplitudes and phases:
\begin{gather}
a_{t}+\frac{1}{2m_{1}}a\chi _{xx}+\frac{1}{m_{1}}a_{x}\chi _{x}=0,  \label{a}
\\
b_{t}+\frac{1}{2m_{2}}b\eta _{xx}+\frac{1}{m_{2}}b_{x}\eta _{x}=0,  \label{b}
\\
-a\chi _{t}+\frac{1}{2m_{1}}a_{xx}-\frac{1}{2m_{1}}a\chi
_{x}^{2}-g_{1}a^{3}-g_{12}b^{2}a=0,  \label{theta} \\
-b\eta _{t}+\frac{1}{2m_{2}}b_{xx}-\frac{1}{2m_{2}}b\eta
_{x}^{2}-g_{2}b^{3}-g_{12}a^{2}b=0.  \label{eta}
\end{gather}

\subsubsection{The analytical approach in the 1D case}

Choosing the constant amplitudes of the two states as $a_{0}$ and $b_{0}$
respectively, we obtain CW (continuous-wave) solutions of the HV\ type of
Eqs. (\ref{a})-(\ref{eta}),
\begin{eqnarray}
&&\hspace{3cm}\chi =-\mu _{1}t+sx\,\,\,,\,\,\,\eta =-\mu _{2}t-sx,
\label{00HV} \\
&&\mu _{1}=g_{1}a_{0}^{2}+g_{12}b_{0}^{2}+\left( s^{2}/2m_{1}\right)
\,\,\,,\,\,\,\mu _{2}=g_{2}b_{0}^{2}+g_{12}a_{0}^{2}+\left(
s^{2}/2m_{2}\right) \;.  \label{mumuHV}
\end{eqnarray}%
Here integer $s$ determines the opposite vorticities in the two components,
without introducing net phase circulation. To address the important issue of
the stability of the HV-CW state, or the zero-vorticity one in the case of $%
s=0$, perturbed solutions to Eqs. (\ref{a})-(\ref{eta}) are looked for as%
\begin{eqnarray}
a\left( x,t\right) &=&a_{0}+\delta a\exp \left( \sigma t+ipx\right) ,  \notag
\\
b\left( x,t\right) &=&b_{0}+\delta b\exp \left( \sigma t+ipx\right) ,  \notag
\\
&&  \label{pert} \\
\chi \left( x,t\right) &=&-\mu _{1}t+sx+\delta \chi \exp \left( \sigma
t+ipx\right) ,  \notag \\
\eta \left( x,t\right) &=&-\mu _{2}t-sx+\delta \eta \exp \left( \sigma
t+ipx\right) ,  \notag
\end{eqnarray}%
where $\sigma $ is the instability growth rate (which may be complex), $p$
is a real wavenumber of the perturbations, while $\delta a,\delta b$ and $%
\delta \chi ,\delta \eta $ are their infinitely small amplitudes. The
substitution of these expressions in Eqs. (\ref{a}) - (\ref{eta}) and
linearization [i.e., the derivation of the respective Bogoliubov - de Gennes
(BdG) equations] yields the following dispersion equation for $\sigma (p)$:%
\begin{equation}
\left\vert
\begin{array}{cccc}
\sigma +i\frac{s}{m_{1}}p & 0 & -\frac{p^{2}}{2m_{1}} & 0 \\
0 & \sigma -i\frac{s}{m_{1}}p & 0 & -\frac{p^{2}}{2m_{2}} \\
-\frac{p^{2}}{2m_{1}}-2g_{1}a_{0}^{2} & -2g_{12}a_{0}b_{0} & -\sigma -i\frac{%
s}{m_{1}}p & 0 \\
-2g_{12}a_{0}b_{0} & -\frac{p^{2}}{2m_{2}}-2g_{2}b_{0}^{2} & 0 & -\sigma +i%
\frac{s}{m_{2}}p%
\end{array}%
\right\vert =0.  \label{detHV}
\end{equation}

Next, we consider two separate cases, depending on the value of $s$.

\subsubsection{Zero-vorticity states, $s=0$}

For $s=0$, determinant (\ref{detHV}) defining the stability takes the
explicit form
\begin{gather}
p^{8}+4\left( g_{1}a_{0}^{2}m_{1}+g_{2}b_{0}^{2}m_{2}\right) p^{6}+4\left[
\sigma ^{2}m_{1}^{2}+\sigma
^{2}m_{2}^{2}-4a_{0}^{2}b_{0}^{2}m_{1}m_{2}\left(
g_{12}^{2}-g_{1}g_{2}\right) \right] p^{4}  \notag \\
+16\left( g_{1}a_{0}^{2}m_{2}+g_{2}b_{0}^{2}m_{1}\right) m_{1}m_{2}\sigma
^{2}p^{2}+16\sigma ^{4}m_{1}^{2}m_{2}^{2}\allowbreak =0.  \label{disp}
\end{gather}%
Due to the periodic b.c. %[see Eq. (\ref{r})],
set by the ring geometry, $p$ is quantized,
\begin{equation}
p=n/r,~n=0\,\,\,,\,\,\,\pm 1,\pm 2,~...  \label{discr}
\end{equation}%
(recall we here fix $r=1$ by means of scaling). The onset of the transition
to the immiscibility (i.e., instability against the phase separation) is
signalled by condition %
%\begin{equation}
$\sigma \left( p=\pm 1/r\right) =0$. It follows from %\end{equation}%
Eq. (\ref{disp}) that this instability takes place at%
\begin{equation}
g_{12}^{2}-g_{1}g_{2}>\left( g_{12}^{2}-g_{1}g_{2}\right) _{\mathrm{cr}%
}\equiv \frac{r^{-2}\left[ 4\left(
m_{1}g_{1}a_{0}^{2}+m_{2}g_{2}b_{0}^{2}\right) +r^{-2}\right] }{%
16m_{1}m_{2}a_{0}^{2}b_{0}^{2}}.  \label{cr}
\end{equation}%
Note that even in the case of $g_{1}=g_{2}=0$ (no self-repulsion), Eq. (\ref%
{cr}) yields a finite threshold for the onset of the phase-separation
instability:%
\begin{equation}
\left( g_{12}^{2}-g_{1}g_{2}\right) _{\mathrm{cr}}|_{g_{1}=g_{2}=0}=\frac{%
r^{-4}}{16m_{1}m_{2}a_{0}^{2}b_{0}^{2}}.  \label{g1=g2=0}
\end{equation}%
This result explicitly demonstrates that periodic b.c. provide for partial
stabilization of the mixed state, in comparison with the infinite free
space, cf. Ref. \cite{Mineev_1974}.

\subsubsection{Hidden-vorticity (HV) states, $s\geq 1$}

As defined above, HV states carry opposite angular momenta in the two
components of the mixture, while the total momentum is zero. In an explicit
form, the corresponding equation (\ref{detHV}), which determines their
stability, takes a very cumbersome form. It becomes relatively simple in the
case of full symmetry in Eqs. (\ref{a})-(\ref{eta}) and (\ref{00HV})-(\ref%
{mumuHV}), \textit{viz}.,
\begin{equation}
m_{1}=m_{2}\equiv m,~g_{1}=g_{2}\equiv g,~g_{12}\equiv 1,~a_{0}=b_{0},
\label{g}
\end{equation}%
for which we obtain $\mu _{1}=\mu _{2}=-\left( g+1\right)
a_{0}^{2}+s^{2}/\left( 2m\right) $. Then, Eq. (\ref{detHV}) can be
explicitly written as%
\begin{equation}
16m^{4}\sigma ^{4}+\left( 32s^{2}+32gma_{0}^{2}+8p^{2}\right)
m^{2}p^{2}\sigma ^{2}+16\left[ \left( gma_{0}^{2}-s^{2}\right)
^{2}-m^{2}a_{0}^{4}\right] p^{4}+8\left( gm\allowbreak
a_{0}^{2}-s^{2}\right) p^{6}+p^{8}=0.  \label{HV}
\end{equation}%
Alternatively, the free term in Eq. (\ref{HV}) (the part which does not
contain $\sigma ^{2}$) can be written as%
\begin{equation*}
16\left[ \left( gma_{0}^{2}-s^{2}\right) ^{2}-m^{2}a_{0}^{4}\right]
p^{4}+8\left( gm\allowbreak a_{0}^{2}-s^{2}\right) p^{6}+p^{8}\equiv \left[
4\left( gma_{0}^{2}-s^{2}\right) +p^{4}\right] ^{2}-16m^{2}a_{0}^{4}p^{4}.
\end{equation*}

\begin{figure}[t]
\caption{(Color online) (a) and (b): Eigenvalues produced by Eq. (\protect
\ref{HV2}) for unstable and stable 1D HV (hidden-vorticity) modes,
respectively. The parameters are $(g,\protect\gamma )=(0.1,0.63)$ in (a),
and $(g,\protect\gamma )=(2,0.5)$ in (b). Positive values of $\Sigma $
imply, according to its definition (\protect\ref{Sigma}), the existence of
an unstable eigenvalue, $\protect\sigma =\pm a_{0}^{2}\protect\sqrt{\Sigma }%
. $ Panel (a) shows that $\Sigma $ vanishes at $P=0$ and $P=6.12$, as
predicted by Eq. (\protect\ref{Sigma=0}). }
\label{HV_analyt}\includegraphics[scale=0.4]{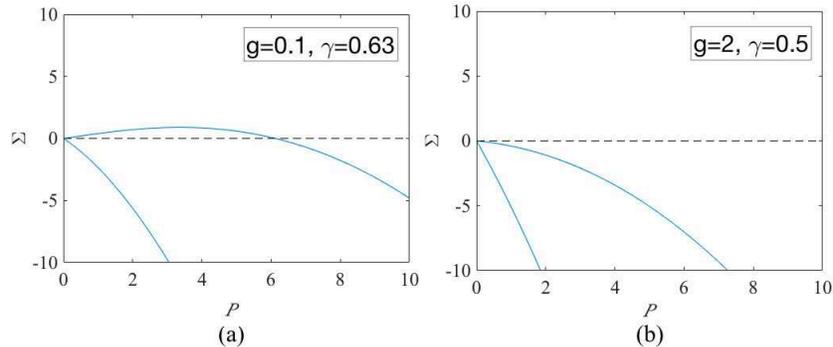}
%\centering{\label{fig111a} %
\end{figure}

Further, Eq. (\ref{HV}) can be cast in the rescaled form,%
\begin{equation}
16\Sigma ^{2}+8\left( 4\gamma +4g+P\right) P\Sigma +16\left[ \left( g-\gamma
\right) ^{2}-1\right] P^{2}+8\left( g-\gamma \right) P^{3}+P^{4}=0.
\label{HV2}
\end{equation}%
with
\begin{equation}
\Sigma \equiv \sigma ^{2}/a_{0}^{4}\,\,,\,\,\,P\equiv p^{2}/\left(
ma_{0}^{2}\right) ,  \label{Sigma}
\end{equation}%
which implies measuring $\sigma ^{2}$ and $p^{2}$ in their natural units,
and demonstrates that the equation depends on two parameters only,
\begin{equation}
\gamma \equiv s^{2}/\left( ma_{0}^{2}\right) \,\,\mathrm{\,and}\,\,\,g.
\label{gamma}
\end{equation}%
Typical examples of the $\Sigma (P)$ dependence for unstable and stable HV
states, produced by Eq. (\ref{HV2}), are presented in Figs.\ref{HV_analyt}%
(a) and (b), respectively. Note that the underlying quantization condition (%
\ref{discr}) implies that $P$ takes, in fact, only discrete values:%
\begin{equation}
P_{n}=\frac{n^{2}}{ma_{0}^{2}r^{2}}\,\,\,,\,\,\,b=0,1,2,....  \label{P}
\end{equation}

The stability condition means that, for all discrete values of $P$, given by
Eq. (\ref{P}), Eq. (\ref{HV2}) must produce negative real solutions for $%
\Sigma $. Full consideration of the stability conditions following from Eq. (%
\ref{HV2}) is too cumbersome for the analytical investigation. Nevertheless,
for the infinite system [$r\rightarrow \infty $, i.e., considering $P$ as a
continuous variable, rather than the discrete one, defined by Eq. (\ref{P}%
)], it is easy to obtain the stability condition in the limit of $%
P\rightarrow 0$, for which Eq. (\ref{HV2}) amounts to%
\begin{equation}
16\Sigma ^{2}+8\left( 4\gamma +4g\right) P\Sigma +16\left[ \left( g-\gamma
\right) ^{2}-1\right] P^{2}=0.  \label{smallP}
\end{equation}%
It is easy to see that Eq. (\ref{smallP}) produces stable solutions, i.e.,
real $\Sigma <0$ [see Eqs. (\ref{pert}) and (\ref{Sigma})], under condition $%
\left\vert g-\gamma \right\vert \geq 1$, i.e., in either of the two cases:
\begin{equation}
g\geq 1+\gamma \,,\mathrm{~or}\,\,\,\gamma \geq 1+g\,.  \label{g>}
\end{equation}%
According to Eq. (\ref{gamma}), conditions (\ref{g>}) hold in the case of a
relatively high nonlinearity (large $g$, or the atom number), or large
hidden vorticity, $s^{2}$, which appears in Eq. (\ref{gamma}), see further
details below.

Further, it is possible to find values of $P$ at which $\Sigma $ vanishes:
substituting $\Sigma =0$ in Eq. (\ref{HV2}), one obtains
\begin{equation}
P=0\,\,\,\mathrm{and}\,\,\,P=4\left( \gamma -g\pm 1\right) .  \label{Sigma=0}
\end{equation}%
If Eq. (\ref{Sigma=0}) yields $P\leq 0$, i.e., $g\geq 1+\gamma $, cf. Eq. (%
\ref{g>}), this implies that the HV states are \emph{completely stable} both
for the infinite system and the ring (since, by definition, $P$ may only be
positive).
% this, in the combination with Eq. (\ref{g>}) or (\ref{gamma>}) means that the HV is completely stable. Finally, it is easy to see that Eq. (\ref{g>}) the condition gives the condition of the complete stability, both the infinite system and the ring alike.

Note the modulational stability of uniform HV states with periodic boundary
conditions was studied in Ref. \cite{brtka_gammal_2010} for the case of the
attractive nonlinearity (on the contrary to the repulsive nonlinearity
considered here), for which it was found that the HV-CW states can \emph{%
never} be stable.

\subsection{The two-dimensional setting}

Stationary solutions to Eq. (\ref{phipsi}) are looked for in the general
form:
\begin{equation}
\phi \left( r,\theta ,t\right) =\Phi _{S}\left( r\right) \exp \left( -i\mu
_{1}t+iS_{1}\theta \right) \,\,\,,\,\,\,\psi \left( r,\theta ,t\right) =\Psi
_{S}\left( r\right) \exp \left( -i\mu _{2}t+iS_{2}\theta \right) ,
\label{StaSol2d}
\end{equation}%
where $(r,\theta )$ are the polar coordinates, $\mu _{1,2}$ chemical
potentials of the two components, $S_{1,2}=0,1,2,...$ their vorticities \cite%
{old}, and real wave functions $\Phi $ and $\Psi $ obey the radial equations:

\begin{eqnarray}
\mu _{1}\Phi _{S}+\frac{1}{2m_{1}}\left( \frac{d^{2}}{d^{2}r}+\frac{1}{r}%
\frac{d}{dr}-\frac{S_{1}^{2}}{r^{2}}\right) \Phi _{S}-\left( g_{1}\left\vert
\Phi _{S}\right\vert ^{2}+g_{12}|\Psi _{S}|^{2}\right) \Phi _{S} &=&0,
\notag \\
&&  \label{StaEq2d} \\
\mu _{2}\Psi _{S}+\frac{1}{2m_{2}}\left( \frac{d^{2}}{d^{2}r}+\frac{1}{r}%
\frac{d}{dr}-\frac{S_{2}^{2}}{r^{2}}\right) \Psi _{S}-\left( g_{2}\left\vert
\Psi _{S}\right\vert ^{2}+g_{12}|\Phi _{S}|^{2}\right) \Psi _{S} &=&0,
\notag
\end{eqnarray}%
To address its stability, we replace the stationary solutions with perturbed
ones:
\begin{eqnarray}
\phi \left( r,\theta ,t\right) &=&\left( \Phi \left( r\right)
+u_{1}e^{\sigma t+il\theta }+u_{2}^{\ast }e^{\sigma ^{\ast }t-il\theta
}\right) e^{-i\mu _{1}t+iS_{1}\theta },  \notag \\
&&  \label{PtbSol} \\
\psi \left( r,\theta ,t\right) &=&\left( \Psi \left( r\right)
+v_{1}e^{\sigma t+il\theta }+v_{2}^{\ast }e^{\sigma ^{\ast }t-il\theta
}\right) e^{-i\mu _{2}t+iS_{2}\theta },  \notag
\end{eqnarray}%
where $l$ is an integer azimuthal index of the perturbation with components $%
u_{1,2}$, $v_{1,2}$, and $\sigma $ is the instability growth rate.

Linearization around the stationary solutions leads to the BdG equations for
the two-component condensate:
\begin{eqnarray}
-\frac{1}{2}\left(u_{1}^{\prime \prime }+\frac{1}{r}u_{1}^{\prime }-\frac{%
(S+l)^{2}}{r^{2}}u_{1}\right)+g_{1}\Phi ^{2}(2u_{1}+u_{2})+g_{12}\Psi
^{2}u_{1}+g_{12}\Psi \Phi (v_{1}+v_{2})-\mu _{1}u_{1} &=&i\sigma u_{1},
\notag \\
-\frac{1}{2}\left(u_{2}^{\prime \prime }+\frac{1}{r}u_{2}^{\prime }-\frac{%
(S-l)^{2}}{r^{2}}u_{2}\right)+g_{1}\Phi ^{2}(2u_{2}+u_{1})+g_{12}\Psi
^{2}u_{2}+g_{12}\Psi \Phi (v_{1}+v_{2})-\mu _{1}u_{2} &=&-i\sigma u_{2},
\notag \\
-\frac{1}{2}\left(v_{1}^{\prime \prime }+\frac{1}{r}v_{1}^{\prime }-\frac{%
(S+l)^{2}}{r^{2}}v_{1}\right)+g_{2}\Psi ^{2}(2v_{1}+v_{2})+g_{12}\Phi
^{2}u_{1}+g_{12}\Psi \Phi (u_{1}+u_{2})-\mu _{2}v_{1} &=&i\sigma v_{1},
\notag \\
-\frac{1}{2}\left(v_{2}^{\prime \prime }+\frac{1}{r}v_{2}^{\prime }-\frac{%
(S-l)^{2}}{r^{2}}v_{2}\right)+g_{2}\Psi ^{2}(2v_{2}+v_{1})+g_{12}\Phi
^{2}u_{2}+g_{12}\Psi \Phi (u_{1}+u_{2})-\mu _{2}v_{2} &=&-i\sigma v_{2},
\label{eigfunct}
\end{eqnarray}%
where the prime stands for $d/dr$. Instabilities are predicted when
numerical solution of Eq. (\ref{eigfunct}) produces eigenvalues with Re$%
(\sigma )\neq 0$. In the 1D version of Eq. (\ref{eigfunct}), $d^{2}/dr^{2}$
is replaced by $d^{2}/dx^{2}$, and terms $\sim 1/r$ and $1/r^{2}$ are absent.

Previously, BdG equations were addressed in the annular geometry defined not
by the rigid boundaries, as per Eq. (\ref{r}), but by weak confinement
constructed as the sum of Gaussian and harmonic oscillator potentials \cite%
{abad_sartori_2014}. BdG equations for two-component condensates were also
studied in other settings, including free space \cite{Shungo_2011,CKLaw_2001}%
, 1D configurations \cite{shimodaira_kishimoto_2010}, and a full analytical
solution \cite{anoshkin_wu_2013}.

\begin{figure}[t]
\centering{\label{fig1a} \includegraphics[scale=0.16]{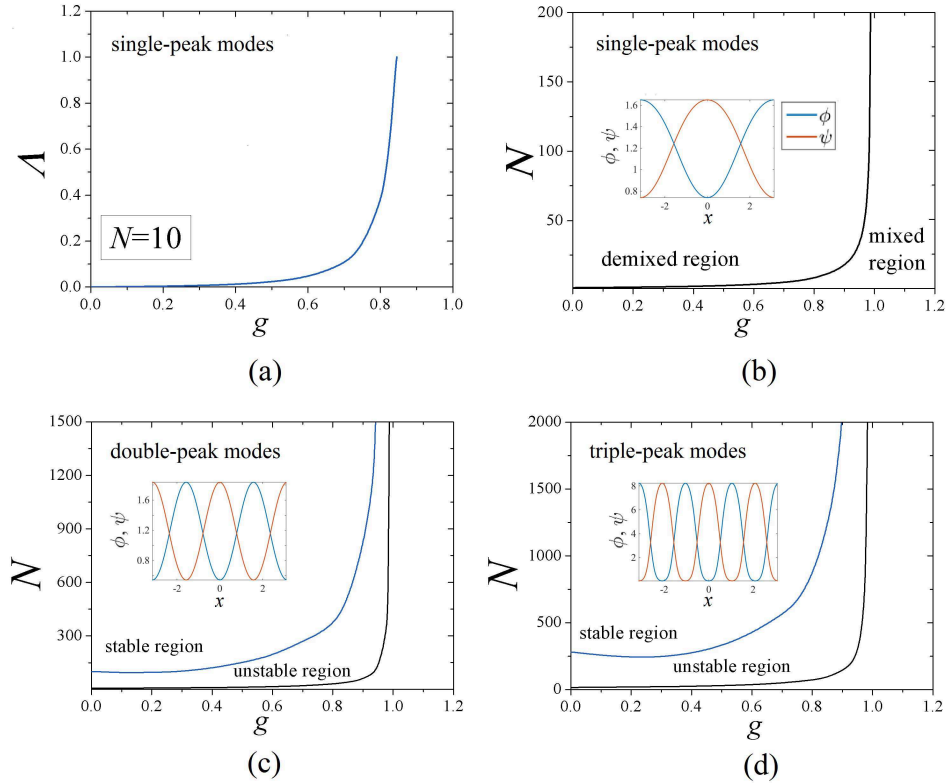}}
\caption{(Color online) Numerically generated existence and stability areas
for 1D mixed and demixed modes: (a) Overlap integral $\Lambda $ of the $%
\protect\phi $ and $\protect\psi $ components, defined as per Eq. (\protect
\ref{Lambda}), versus self-repulsion coefficient $g$, for single-peak modes
with fixed norm $N=10$, see Eq. (\protect\ref{N}); recall that the
inter-species repulsion coefficient is fixed to be $g_{12}=1$. Stability and
existence areas for the single-peak, double-peaks, and triple-peak modes in
the plane of $(g,N)$ are displayed, respectively, in panels (b), (c), and
(d). Black bottom curves in (b), (c) and (d) separate demixed (left) and
mixed (right) states, while blue curves in (c) and (d) separate stable and
unstable demixed ones. Insets in panels (b), (c) and (d) represent,
respectively, typical examples of a stable single-peak mode [with parameters
$(N,g,\Lambda )=(10,0.8,0.37)$], unstable double-peak one [for $(N,g,\Lambda
)=(10,0.1,0.14)$], and unstable triple-peak state, for $(N,g,\Lambda
)=(150,0.1,6.34\times 10^{-4})$. }
\label{1d_N_g}
\end{figure}

\section{Results and discussion}

\subsection{The one-dimensional regime}

Stationary solutions to Eqs.(\ref{phipsi}) were produced numerically by
means of the imaginary-time-evolution method, using different inputs. Then,
stability of these solutions was identified through the calculation of their
eigenvalue spectra, using the 1D version of Eq. (\ref{eigfunct}), and
further verified by direct numerical simulations of the perturbed evolution.
The system conserves the total norm, i.e., scaled number of atoms. $N_{%
\mathrm{total}}=N_{\phi }+N_{\psi }\equiv \int_{0}^{2\pi }\left( |\phi
|^{2}+|\psi |^{2} \right) dx$. Below, we report numerical results obtained
for the basic symmetric states, with $m_{1}=m_{2}=1$ (fixed by scaling), $%
g_{1}=g_{2}\equiv g$, $g_{12}=1$ [also fixed by scaling, cf. Eq. (\ref{g})]
and equal 1D norms in the two components,
\begin{equation}
\int_{0}^{2\pi }\left\vert \phi (x)\right\vert ^{2}dx=\int_{0}^{2\pi
}\left\vert \psi (x)\right\vert ^{2}dx\equiv N.  \label{N}
\end{equation}

In the miscible phase, the two components of the condensates overlap with
each other, whereas they spatially separate in the immiscible phase. A
measure to characterize these phases is the overlap integral, which we
define here in the 2D form, as it will be used below in the analysis of the
2D setting:
\begin{equation}
\Lambda =\frac{\left[ \int \int dxdy|\phi (x,y)|^{2}|\psi (x,y)|^{2}\right]
^{2}}{\left[ \int \int dxdy|\phi (x,y)|^{4}\right] \left[ \int \int
dxdy|\psi (x,y)|^{4}\right] },  \label{Lambda}
\end{equation}%
the reduction of the definition to the 1D limit being obvious. In this work,
we identify demixed and mixed states as those with $\Lambda \neq 1$ and $%
\Lambda =1$, respectively.

\begin{figure}[t]
\centering{\label{fig2a} \includegraphics[scale=0.4]{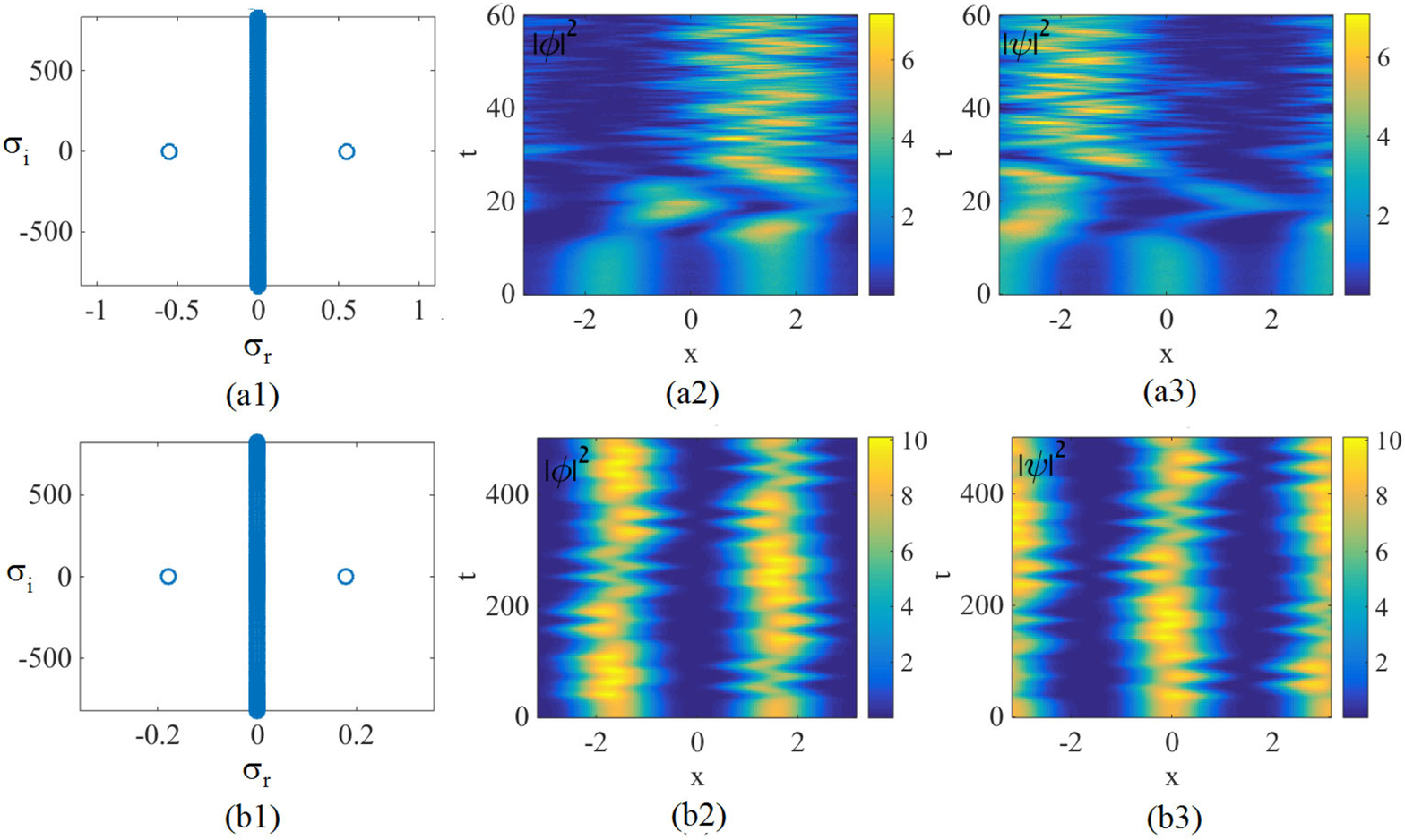}}
\caption{(Color online) Eigenvalue spectra and direct simulations for two
generic types of unstable 1D double-peaks modes. Panels (a1-a3), for the
parameter set $(N,g,\Lambda )=(10,0.1,0.1354)$, taken far from the stability
boundary, show the transformation into a persistent single-peak mode with
irregular oscillations. Panels (b1-b3), for $(N,g,\Lambda )=(20,0.1,0.0104)$%
, show oscillations of the weakly unstable double-peak state around itself
close to the instability boundary.}
\label{1d_instb_2p}
\end{figure}

As expected \cite{Mineev_1974,Lingua}, demixed states exist only when the
cross-repulsion is stronger than the self-repulsion, i.e., $g \leqslant
g_{12}=1$. They are characterized by local density peaks in each component,
located so that a peak in one component coincides with a density minimum in
the other, see insets to Figs. \ref{1d_N_g}(b)-(d). Overlap integral (\ref%
{Lambda}) for 1D single-peak demixed modes is displayed in Fig. \ref{1d_N_g}%
(a), as a function of self-repulsive coefficient $g$, for a fixed norm, $N=10
$. In this figure, the demixed single-peak mode terminates at $g_{\mathrm{cr}%
}=0.845$, only the uniformly mixed state existing at $g>g_{\mathrm{cr}}$.
This numerically identified critical value \emph{exactly coincides with} the
analytical prediction given by Eq. (\ref{cr}). Further, the existence area
for demixed single-peak and mixed modes is presented in Fig. \ref{1d_N_g}%
(b). The boundary between them, analytically predicted by Eq. (\ref{cr}),
also exactly coincides with the numerically found counterpart, shown by the
black curve in Fig. \ref{1d_N_g}(b). The single-peak demixed modes are
completely stable in their existence domain, which is consistent with
earlier findings \cite{white_hennessy_2016}.

Stability and existence areas of demixed double- and triple-peak modes are
displayed in parameter plane $(g,N)$ in Figs. \ref{1d_N_g}(c) and (d),
respectively, and typical examples of such modes are displayed in their
respective insets. An essential finding is that, unlike the single-peak
modes which are always stable, states with two and three peaks feature
instability areas in Figs. \ref{1d_N_g}(c) and (d), being stable only for a
sufficiently large norm.

\begin{figure}[t]
\centering{\label{fig101a} \includegraphics[scale=0.4]{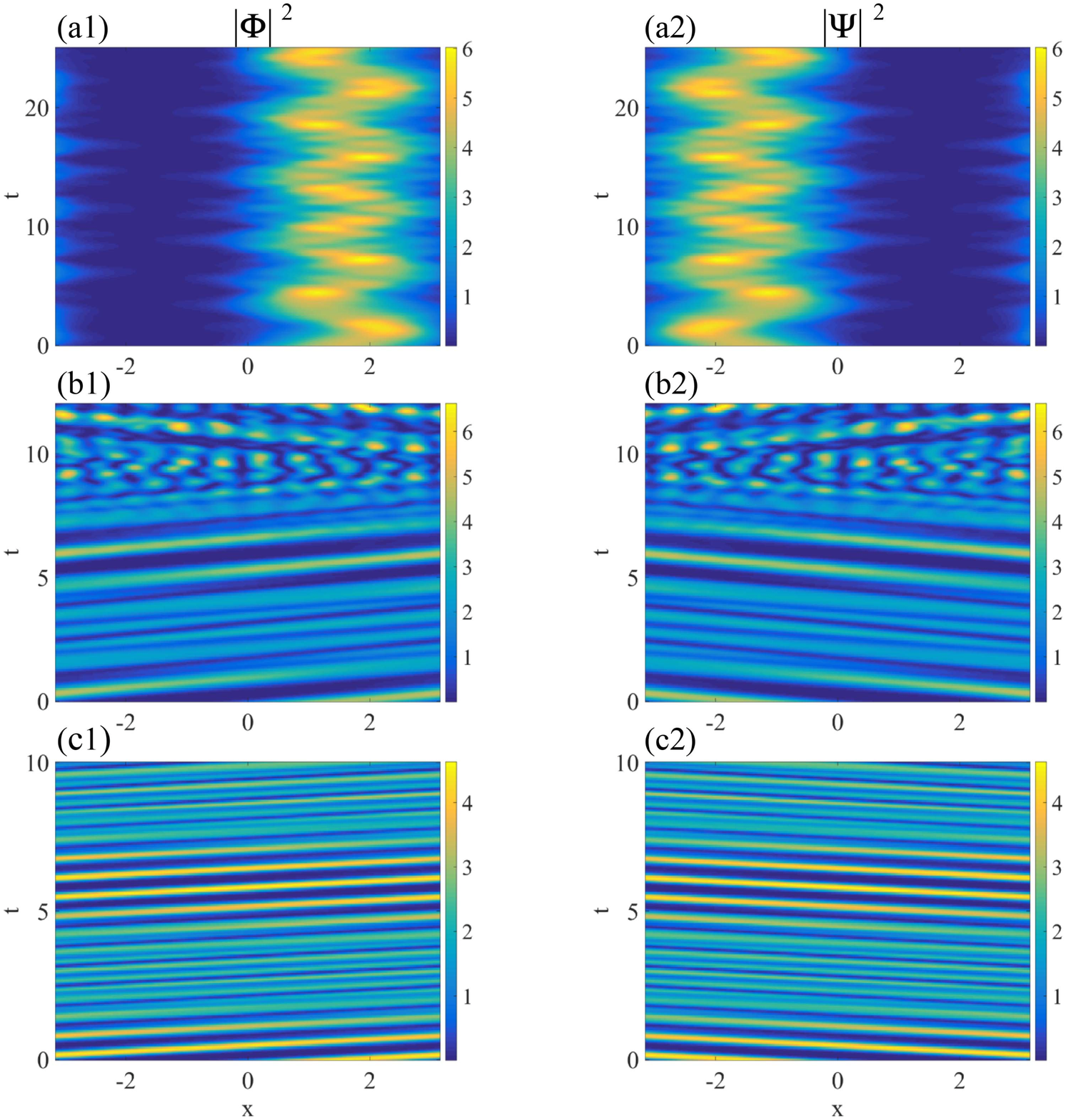}}
\caption{(Color online) Collisions between components of 1D single-peak
modes, with parameters $(N,g,\Lambda )=(10,0.1,1.8 \times 10^{-3})$,
initiated by kick (\protect\ref{kick}) with $k=0.5$ (a1, a2), $k=5$ (b1,
b2), and $k=10$ (c1, c2). }
\label{1peak_kick}
\end{figure}

The numerical analysis reveals two instability scenarios for the double-peak
mode. If it is taken in the area far from the stability boundary in Fig. \ref%
{1d_N_g}(c), the real parts of the corresponding eigenvalues $\sigma $ are
relatively large [see Eq. (\ref{PtbSol})], and the mode spontaneously
transforms into an oscillating single-peak state, see Fig. \ref{1d_instb_2p}%
(a1-a3). If the unstable mode is selected close to the instability boundary,
with smaller real parts of the eigenvalues, it oscillates around itself,
rather than transforming into a single-peak state. Similar to the
double-peak states, unstable triple-peak ones transform into oscillating
single-peak modes far from the corresponding stability boundary, and
persistently oscillate around themselves, if taken close to boundary, see
Fig. \ref{1d_instb_2p}(b1-b3).

We also simulated collision between single-peak demixed components, set in
motion by applying opposite kicks to them:%
\begin{equation}
\phi (x,t=0)=\phi (x)e^{ikx},~\psi (x,t=0)=\psi (x)e^{-ikx}.  \label{kick}
\end{equation}%
Figure \ref{1peak_kick} shows that, for the kick small enough ($k=0.5$), the
peaks in the two components periodically bounce back from each other, which
is accompanied by some randomization of the patterns. On the other hand,
under the action of a strong kick (e.g., $k=5$), the moving components pass
through each other for about five times, but eventually suffer randomization
too, as shown in Fig. \ref{1peak_kick}(b1,b2). Under the action of a still
stronger kick, $k=10$, the components kept passing through each as long as
the simulations were run, see Fig. \ref{1peak_kick}(c1,c2).

\begin{figure}[t]
\centering{\label{fig301a} \includegraphics[scale=0.5]{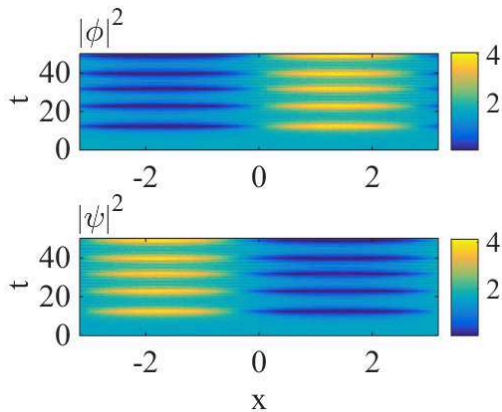}}
\caption{(Color online) Numerically simulated evolution of an unstable 1D
mixed mode, showing periodic transformations between mixed and demixed
states. The parameters are $(N,g,\Lambda )=(10,0.6,1)$. }
\label{DR_mixed_mode}
\end{figure}

It is also relevant to simulate evolution of unstable mixed (uniform)
states, which is displayed in Fig.\ref{DR_mixed_mode}. The instability
triggers periodic transformations between the mixed state and a single-peak
demixed one, with the period $\approx 10$ in this case.

%For experimental consideration, it is quite necessary to compare the energy (defined in Eq.(\ref{1_hamilt})) between different states, see Table. \ref{tab:table1}. With same value of $g$ and $N$, the single-peak one has the lowest energy, while the unitary mixed state has the highest energy among these four states.

%\begin{figure}[tbp]
%\centering{\label{fig201a} \includegraphics[scale=0.4]{201a.eps}}
%\caption{(Color online) (a)Typical examples of unstable 1D triple-peaks mode
%$\protect\phi $ and $\protect\psi $, with parameters set $(N,g,\Lambda )=(150,0.1,0)$%
%. (b)The corresponding eigenvalue spectrum. (c,d)Direct evolution for
%unstable $\protect\phi $ and $\protect\psi $.}
%\label{Example3}
%\end{figure}

%\begin{figure}[tbp]
%\centering{\label{fig3a} \includegraphics[scale=0.17]{3a.eps}}
%\caption{(Color online) Existence and stability areas for 1D mixed and demixed modes:
%(a)Overlap integral $\Lambda $ of $\protect\phi $ and $\protect\psi $,
%versus self-repulsive coefficient $g$, for single-peak modes and its corresponding existence boundary (b).  Stability and existence area of double-peaks modes (c), and triple-peaks modes (d) in panel $(g_{1,2},N)$.
%The down black curves for (b), (c) and (d) separate demixed (left) and mixed (right) solutions. }
%\label{Analyt}
%\end{figure}

\begin{figure}[b]
\centering{\label{fig4a} \includegraphics[scale=0.18]{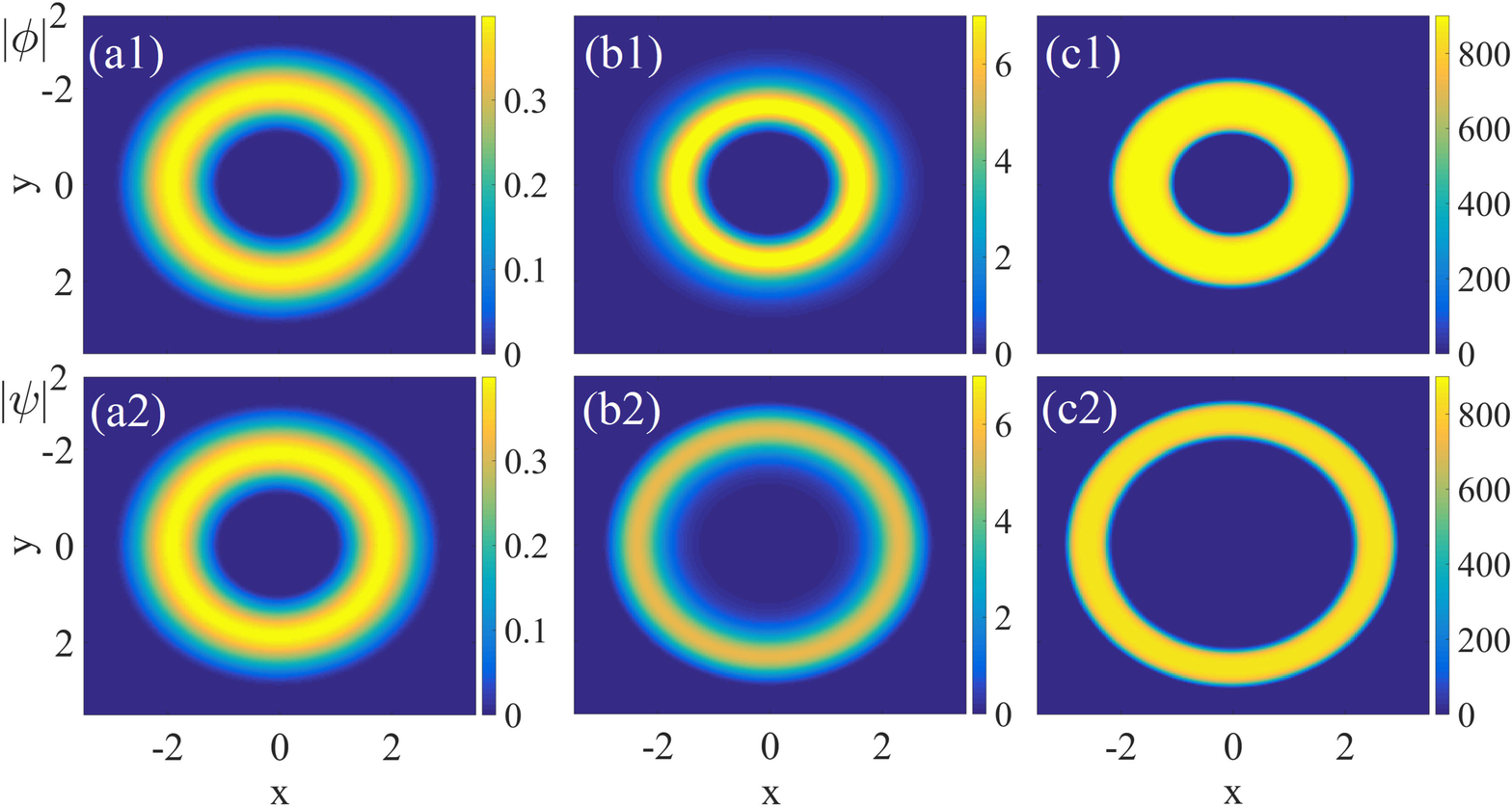}}
\caption{(Color online) Typical examples of 2D zero-vorticity states ($%
S_{1,2}=0$) with $g=0.1$ and width $w=2$. (a1,a2): An unstable mixed state
with $N=5$ and $\Lambda =1$; (b1,b2): an unstable demixed state with $N=60$
and overlap parameter $\Lambda =0.1307$ [see Eq. (\protect\ref{Lambda})];
(c1,c2): a stable strongly demixed state with $N=9000$ and $\Lambda =4.45
\times 10^{-6}$.}
\label{examp2d}
\end{figure}

\subsection{Two-dimensional regime}

Focusing on the phase-separation scenarios, we identify two different types
of 2D demixed modes, namely, those which may be defined as demixed in the
radial direction (cf. Ref. \cite{mason_2013}), and azimuthally demixed ones,
cf. Refs. \cite{mason_2013,abad_sartori_2014,shimodaira_kishimoto_2010}. In
previous works, similar scenarios of the phase separation were also reported
for other binary systems, which include rotation \cite%
{shimodaira_kishimoto_2010,white_hennessy_2016} and spin-orbit coupling \cite%
{white_zhang_2017}.

\subsubsection{Radially-demixed modes}

In the consideration of the 2D setting subject to b.c. (\ref{r}), we focus,
as above, on the scaled symmetric system, with $m_{1}=m_{2}\equiv 1$, $%
g_{1}=g_{2}\equiv 1$, $g_{12}=1$, and $r_{\mathrm{inner}}=1$, cf. Eqs. (\ref%
{g}) and (\ref{inner}). Stability of 2D modes was identified by the
computation of the eigenvalue spectra in the corresponding Bogoliubov-de
Gennes (BdG) equations (\ref{eigfunct}), and further verified by direct
simulations.

\begin{figure}[t]
\centering{\label{fig5a} \includegraphics[scale=0.2]{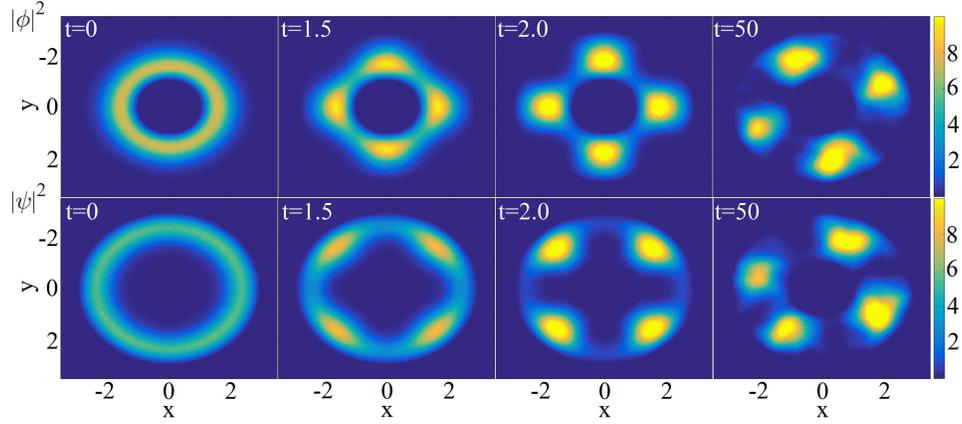}}
\caption{(Color online) Density snapshots of the evolution of an unstable 2D
radially-demixed mode shown in Fig. \protect\ref{examp2d} (middle),
revealing spontaneous formation of azimuthally-demixed states.}
\label{2dinstb}
\end{figure}

First, we address 2D zero-vorticity states, including mixed and
radially-demixed ones, which may be both stable and unstable (at larger and
smaller values of the norm, respectively), as shown in Fig. \ref{examp2d}. A
typical example of the evolution of unstable 2D radially-demixed states with
vorticities $S_{1,2}=0$ is shown in Fig. \ref{2dinstb}. It is observed that
the unstable state spontaneously evolves into an azimuthally-demixed one.

\begin{figure}[b]
\centering{\label{fig6a} \includegraphics[scale=0.5]{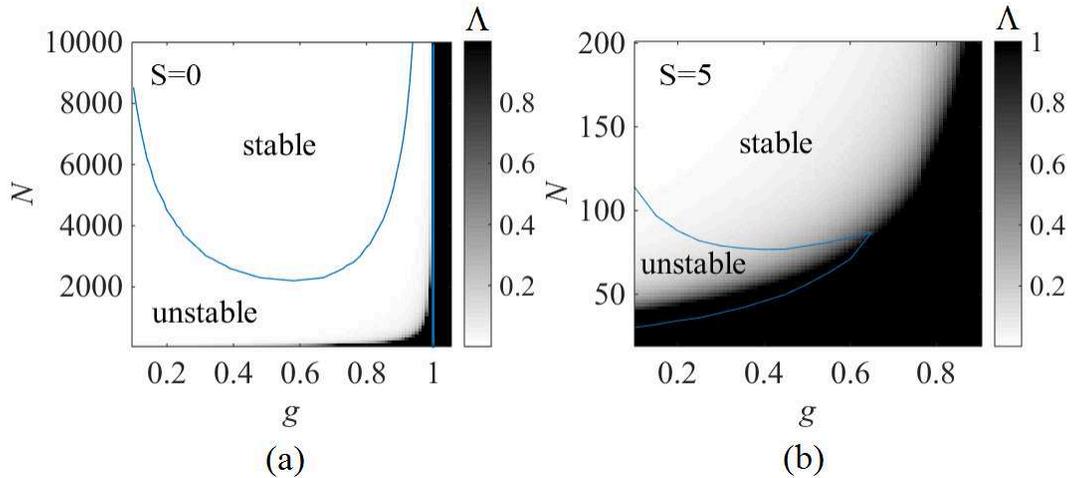}}
\caption{(Color online) Existence and stability areas in the $(g,N)$ plane
for 2D mixed and radially-demixed states, in the annular domain with width $%
w=2$ [see Eq. (\protect\ref{w})]. The overall vorticity is $S=0$ in (b1) and
$S=5$ in (b2). The gray-scaled shading shows the corresponding values of the
overlap parameter $\Lambda $, see Eq. (\protect\ref{Lambda}).}
\label{Lambda_N}
\end{figure}

A noteworthy finding is that the system supports \emph{stable} 2D
radially-demixed states with \emph{arbitrarily high} vorticities $%
S_{1}=S_{2}\equiv S$. We first analyze 2D demixed states with $S=0$ and $S=5$
in the parameter space of $(g,N,\Lambda )$, see Fig. \ref{Lambda_N}. It is
seen that the solutions are stable (similar to what was found above for
other configurations) above a threshold value of the norm, $N>N_{\mathrm{th}%
} $. We stress that the stability threshold is much lower for $S=5$ than for
$S=0$ [note different scales of vertical axes in Figs. \ref{Lambda_N}(a) and
(b)].

\begin{figure}[th]
\centering{\label{fig9a} \includegraphics[scale=0.16]{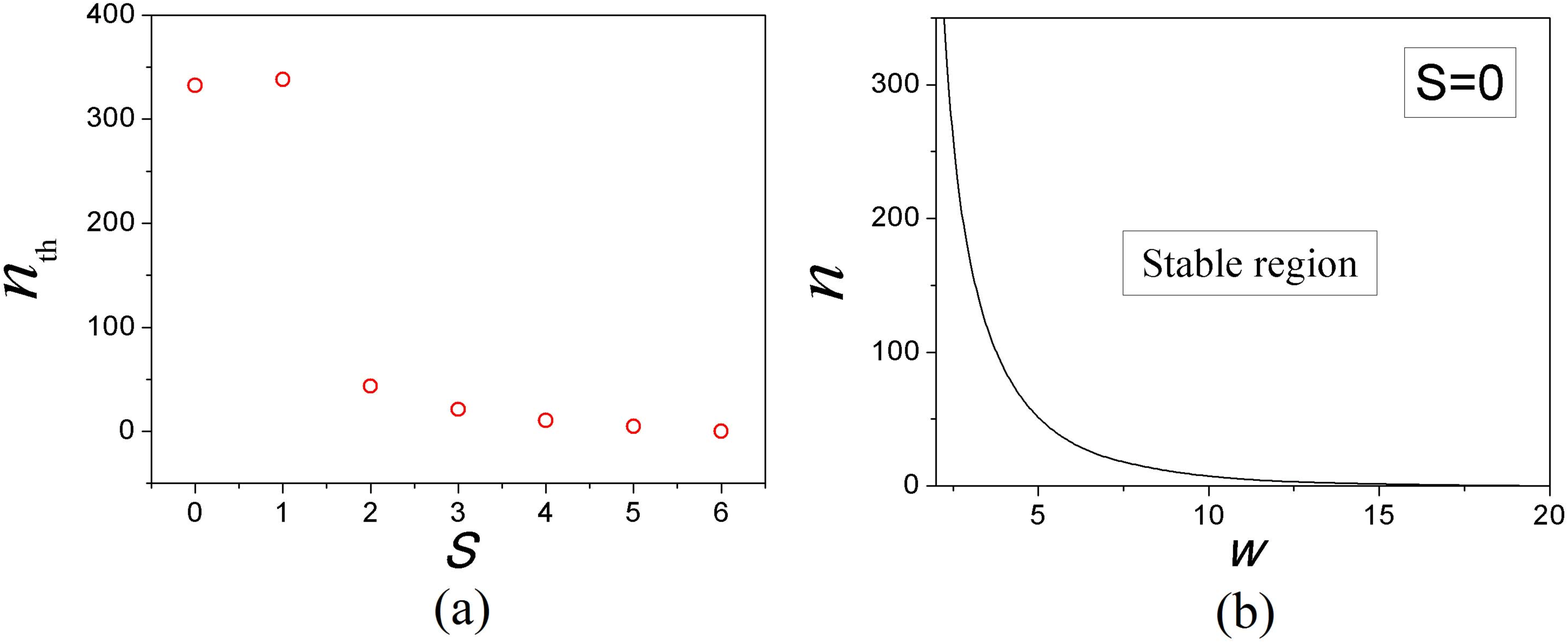}}
\caption{(Color online) (a) The threshold value of density (\protect\ref%
{density}) of 2D radially-demixed states versus their vorticity $S$, the
solutions being stable at $n\geq n_{\mathrm{th}}$. The corresponding
parameter set is $(w,g)=(2,0.1)$, see Eqs. (\protect\ref{g}) and (\protect
\ref{w}). (b) The stability region for the radially-demixed state with $S=0$
in the plane of plane $(n,w)$, for $g=0.1$. The solutions are stable above
the solid curve. }
\label{StbRegionS0}
\end{figure}

To further explore how the vorticity affects the stability of the 2D demixed
states, we define the atomic density,%
\begin{equation}
n=\frac{N}{\pi (r_{\mathrm{outer}}^{2}-1)}  \label{density}
\end{equation}%
[recall that the inner radius of the annulus is scaled to be $1$, as per Eq.
(\ref{inner})], and display the stability-threshold value of $n$ as a
function of $S$ in Fig. \ref{StbRegionS0}(a). A salient feature is the steep
drop of $n_{\mathrm{th}}$ while $S$ increases from $1$ to $2$, which is
followed by gradual decrease of the threshold with further increase of $N$.
Thus, the vorticity helps to strongly stabilize the axially symmetric states
in the annular domain.

It is also relevant to investigate an effect of the annulus' width $w$,
defined as per Eq. (\ref{w}), on the stability. For the zero-vorticity
radially-demixed states, the respective stability diagram in parameter panel
$(n,w)$ is presented in Fig. \ref{StbRegionS0}(b)). It is seen that the
stability area strongly broadens with the increase of $w$, i.e., as it might
be expected, stable radially-demixed modes prefer broad annular domains.

\begin{figure}[b]
\centering{\label{fig7a} \includegraphics[scale=0.15]{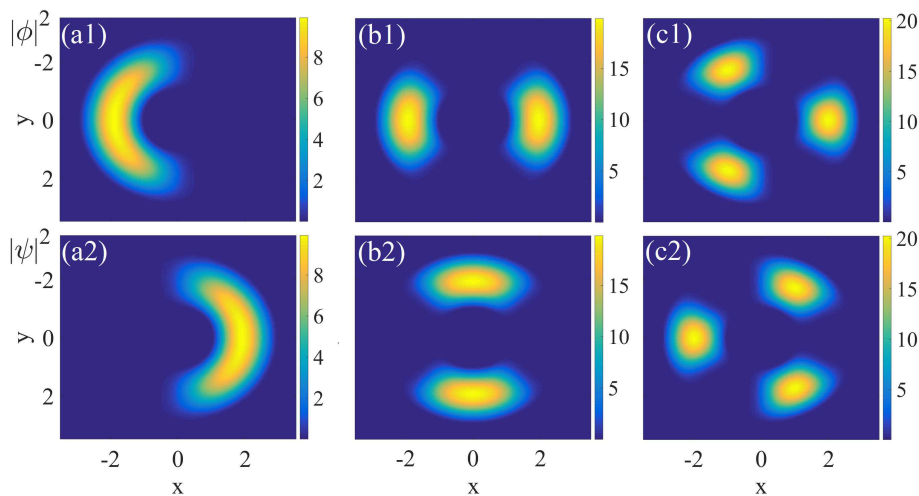}}
\caption{(Color online) Typical examples of 2D azimuthally-demixed modes
with zero vorticity, for $g=0.1$ and the annulus' width $w=2$. (a1,a2) A
stable single-peak mode with total norm $N=50$. (b1,b2) A stable double-peak
mode with $N=100$. (c1,c2) An unstable triple-peak mode with $N=100$.}
\label{quasi2d}
\end{figure}

\begin{figure}[b]
\centering{\label{fig8a} \includegraphics[scale=0.15]{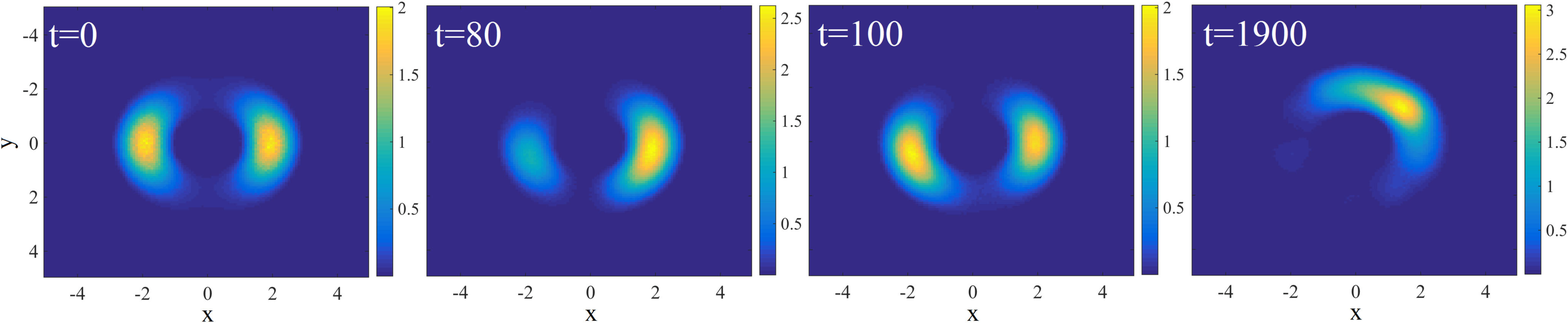}}
\caption{(Color online) Density snapshots of the evolution of an unstable 2D
double-peak azimuthally-demixed mode (only the $|\protect\phi |^{2}$
component is displayed, as the complementary evolution of $|\protect\psi|^2$
is similar) for $(g,w,N)=(0.1,2,20)$. The unstable mode spontaneously
transforms into a stable single-peak one.}
\label{dr_2d_angualr}
\end{figure}

\subsubsection{Azimuthally-demixed modes (with $S=0$)}

The 2D setting supports, as well, stable modes which are phase-separated in
the azimuthal direction (with zero vorticity) \cite{white_hennessy_2016,
white_zhang_2017}, an example of such modes can be seen in Fig. \ref{quasi2d}%
. These modes are related to their 1D counterparts displayed above in the
insets of Figs. \ref{1d_N_g}(b,c,d), and unstable 2D radially-demixed modes
transform into them (in an excited oscillating state), see Fig. \ref{2dinstb}%
.

To illustrate the evolution of those 2D azimuthally-demixed states which are
unstable, we display the evolution of an unstable double-peak state with a
small total norm, $N=20$ in Fig.\ref{dr_2d_angualr} (similar to other states
considered here, they tend to be unstable for relatively small values of $N$%
). It first evolves into a pattern with unequal heights of two peaks, and
then restored the original configuration with equal peaks. After several
cycles of such shape oscillations, it finally settles into an oscillating
single-peak state. The same happens with unstable triple-peak 2D states.
This kind of dynamics resembles what was observed above for unstable double-
and triple-peak states in the 1D geometry, see Fig. \ref{1d_instb_2p}.

On the other hand, we have not found any azimuthally-demixed states with
nonzero vorticity.

%we assume an additional rotation field to rotate the modes. It can be naturally induced by replacing $\theta \rightarrow \theta + \omega t$. One interesting finding is that under the rotation of the field with velocity $\omega=10$, the double-peaks demixed state first evolves into a quadruple-peaks ones, and then evolves back into a double-peaks states. After
%taking this periodic switch for several rounds, it finally evolves into an chaotic state.
%For the case of total norm $N=50$, the critical value of the angular speed which does not destroy the double-peak states is about $\omega=0.75$.
%Stability of these modes is
%analyzed in the $(N,\mu )$ plane, see Fig. \ref{2d_angl_im_mu_N}. If we
%compare it with Fig. \ref{Lambda_N}(b), we can find that the angular-demixed
%modes are more stable than the radially-demixed ones.

%\begin{figure}[t!]
%\centering{\label{fig802a} \includegraphics[scale=0.35]{802a.eps}}
%\caption{(Color online) Typical examples of stable 2D radial-demixed modes
%with $r_{inner}=0$ and parameter set $(g,w,N,S_{1,2})=(0.1,3,12000,0)$ : amplitudes cross-section along $y=0$.
%{\em  I DONT THIKN THIS ADDS ANYHTING TO THE PAPER? ISNT' THIS WELL-KNOWN??? ALSO $r_0$ 	IS NOT DEFINED ANYWHERE IN OUR WORK?}}
%\label{2d_r0}
%\end{figure}

Finally, it makes sense to address demixed modes in the full circle, with $%
r_{\mathrm{inner}}=0$, instead of the annulus, cf. Eq. (\ref{inner}). It has
been found that radially-demixed modes may (quite naturally) exist in the
latter case, while no azimuthally-demixed states were found.

\begin{figure}[b]
\centering{\label{fig11a} \includegraphics[scale=0.35]{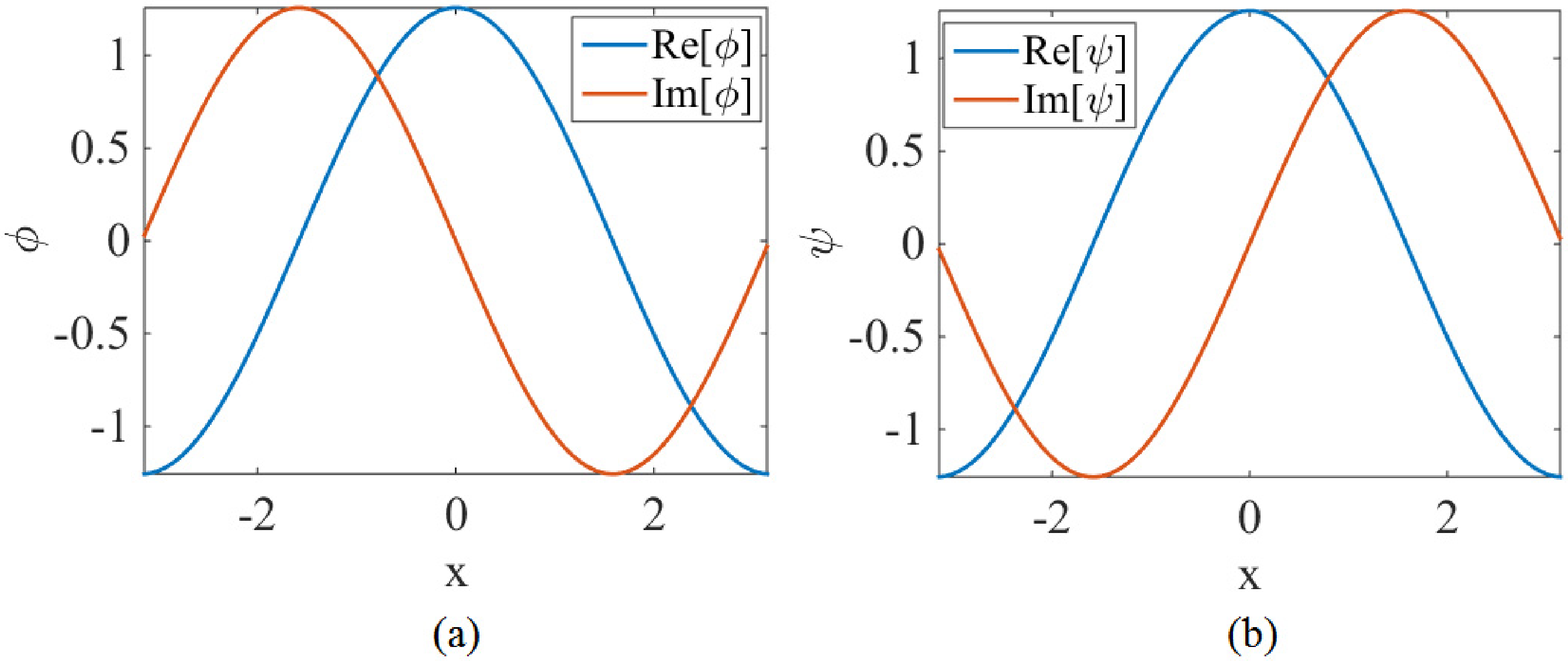}}
\caption{(Color online) A typical example of an unstable one-dimensional HV
(hidden-vorticity) mode with $(g,N,S_{1},S_{2})=(0.1,10,-1,1)$}
\label{HV_example_1d}
\end{figure}

\subsection{Hidden-vorticity states}

\subsubsection{The 1D setting}

A typical example of the HV state, predicted by analytical solution (\ref%
{00HV}), is presented in Fig. \ref{HV_example_1d}. This particular HV state
is an unstable one, as illustrated in Fig. \ref{HV_analyt}(a) by the
dependence of its instability growth rate on the perturbation wavenumber,
which is predicted by Eq. (\ref{HV2}); for comparison, Fig. \ref{HV_analyt}
(b) exhibits the same analytical result for a stable HV state. Simulations
demonstrate that the evolution transforms unstable 1D HVs into stable
single-peak demixed states, with some intrinsic oscillations (not shown here
in detail).

\subsubsection{The 2D setting}

\begin{figure}[t]
\centering{\label{fig12a} \includegraphics[scale=0.17]{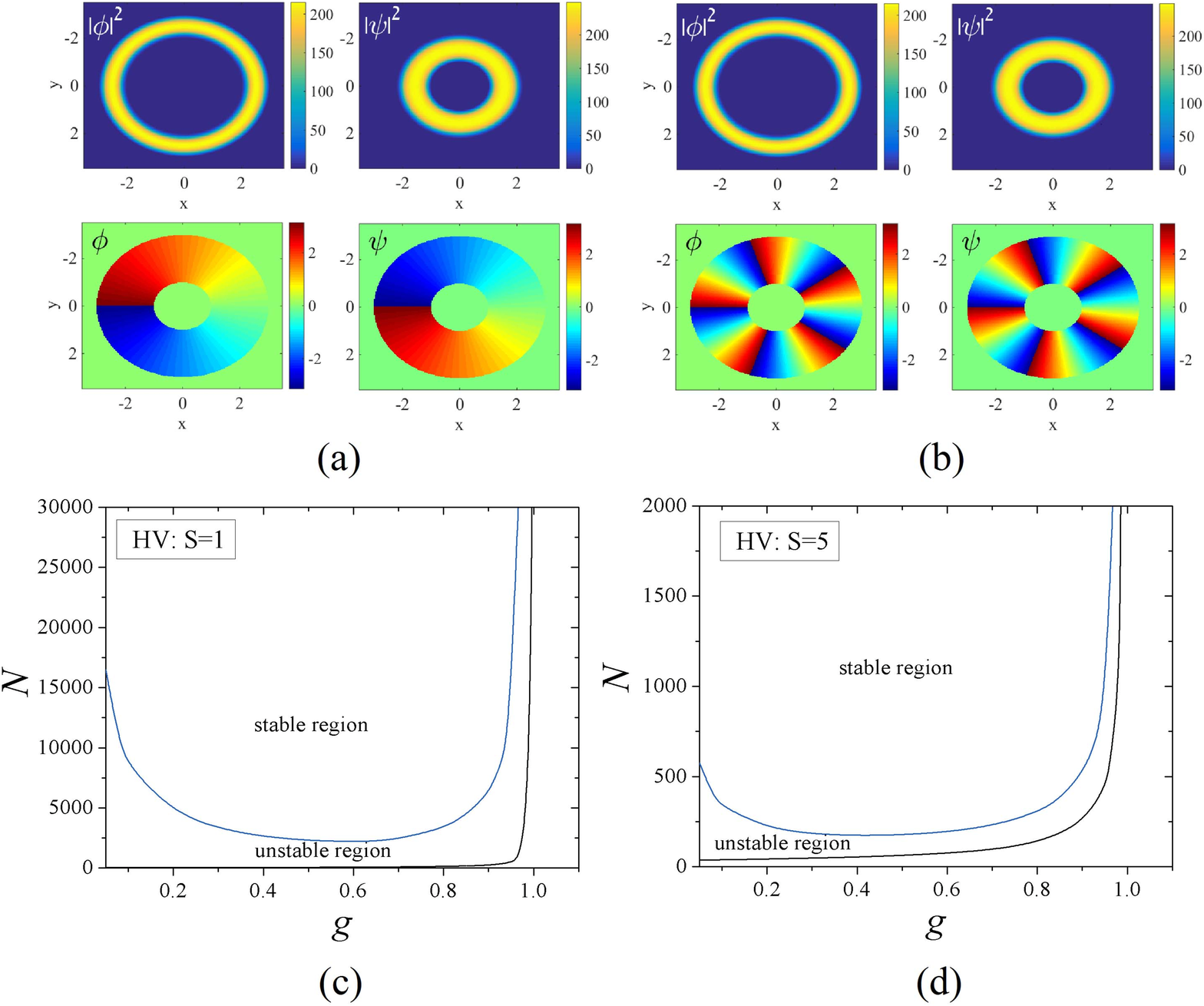}}
\caption{(Color online) Typical examples (the density distribution and phase
structure) of stable 2D radially-demixed HV states: (a) $S_{1,2}=\pm 1$; (b)
$S_{1,2}=\pm 5$. Both examples correspond to the same parameter set, $%
(g_{1,2},w\equiv r_{\mathrm{outer}}-1,N)=(0.1,2,2000)$. Panels (c) and (d)
summarize properties of the respective HV states in parameter plane $\left(
g,N\right) $. In (c) and (d), black curves separate demixed and mixed states
(left and right areas, respectively), while blue curves are stability
boundaries for demixed states.}
\label{2dHV_N_g}
\end{figure}

We have also numerically produced 2D radially-demixed HV states, example of
which, with $S_{1,2}=\pm 1$ and $\pm 5$, are displayed in Figs. \ref%
{2dHV_N_g}(a) and (b), respectively. The same setting may also support 2D
mixed HV states, which we do not consider here in detail, as the demixed
states seem more interesting. Results for the stability of the 2D
radially-demixed HV modes with the same values of $S_{1,2}$ are summarized
in Figs. \ref{2dHV_N_g}(c) and (d). An obviously interesting conclusion
following from the latter plots is that the increase of the hidden
vorticity, $\left\vert S_{1,2}\right\vert $, leads to \emph{stabilization}
of the the HV states [note that difference in the scales of vertical axes in
panels (c) and (d)].

Finally, comparing the total energy of different 2D mixed and demixed states
[see Eq. (\ref{1_hamilt})], which share equal values of the total norm and
angular momentum, we have concluded that the single-peak azimuthally-demixed
states realize the lowest energy, i.e., the system's ground state, while the
totally mixed configuration has the highest energy.

\subsection{Physical Estimates}

To translate the scaled units into the physical ones, we consider the binary
condensate of $^{87}$Rb atoms in two different spin states, such as ones
with $F=1,m_F=1$ and $F=1,m_F=0$, and use the same parameters as experiments
performed with the two-components condensate in a ring \cite%
{beattie_moulder_2013}, with the radius $\lesssim 12$ $\mu $m, and the
scattering length $a_{s}\sim 10$ nm \cite{Egorov_2013}. We conclude that the
stable effectively 1D modes predicted by the present analysis may have the
actual transverse thickness $\sim 3$ $\mu $m, containing up to $\sim 10^{4}$
atoms, while the stable 2D modes, predicted for the same outer radius, $%
\lesssim 12$ $\mu $m, and the inner one $\lesssim 4$ $\mu $m, contain $%
10^{4}\sim 10^{5}$ atoms.

\section{Conclusion}

We have studied the stability and phase diagram of the two-component BEC
loaded in the 2D annular potential box, as well as its 1D limit form
corresponding to a ring. The system was analyzed in the framework of the
mean-field approximation, based on coupled Gross-Pitaevskii equations with
repulsive intra-species and inter-species interactions.
%This system can be realized in binary BECs with an ideal annular optical beam acting on it.

%Family of both 1D and 2D mixed and demixed states are found by means of imaginary-time evolutions method. Their stability are investigated by computation of eigenvalue spectrum for small perturbation, and further verified by means of direct evolution.
In the 1D setting, the demixed (phase-separated) states are identified as
single-, double- and triple-peak modes, with density peaks in one component
coinciding with density minima in the other one. The 1D single-peak demixed
states are all stable, while the double- and triple-peak ones are stable
only above critical values of the total norm, $N$. The unstable double- and
triple-peak modes oscillate around themselves when they are located close to
the instability boundary, or spontaneously transform into stable single-peak
states deeper in the unstable domain of the parameter space. Collisions
between two components of stable demixed single-peak states were studied
too, by applying opposite kicks to the components. The simulations
demonstrate that the weakly kicked components repeatedly bounce from each
other, suffering gradual chaotization, while fast ones pass through each
other. If the kicks are moderately strong, the components originally pass
through each other, and then evolve into the bouncing regime. The evolution
of unstable 1D mixed (spatially uniform) modes shows periodic transitions
between the mixed state and single-peak demixed ones.

In the 2D setting, we have found both radially- and azimuthally-demixed
states, with unstable radially-demixed ones found to evolve into their
azimuthally-demixed counterparts. An essential finding is that the system
supports radially-demixed modes with arbitrarily large overall vorticity $S$%
, which are stable above the threshold value of the norm, $N_{\mathrm{th}}$.
The increase of $S$ leads to stabilization of the modes (decrease of $N_{%
\mathrm{th}}$), with a dramatic drop, following the transition from $S=1$ to
$S=2$, in Fig. \ref{StbRegionS0}(a). The stability area gradually broadens
with the increasing of the annulus' width, $w$, in Fig. \ref{StbRegionS0}%
(b). Similar to the 1D demixed states, 2D azimuthally-demixed ones are also
identified as single-, double- and triple-peak modes. Unstable 2D double-
and triple-peak azimuthally-demixed states (those with relatively small
norms) evolve into oscillating single-peak modes. In the solid circle, taken
instead of the annulus, only radially-demixed modes are found.

%Rotation dynamics of double-peak angular-demixed states revealed a threshold value of rotation speed, above which the original states would be destructed.

Lastly, both 1D and 2D HV (hidden-vorticity) states, with opposite
vorticities in the two components, have been addressed too. The stability
region for 1D HV modes was found analytically, and fully confirmed by the
numerical analysis. Unstable 1D HV modes with components vorticities $%
S_{1,2}=\pm 1$ showed evolve into oscillating single-peak demixed modes. The
stability domain for 2D radially-demixed HV modes expands with the increase
of the hidden vorticity, $\left\vert S_{1,2}\right\vert $.

\section*{Acknowledgments}

This work was supported, in part, by grants No.11874112 and No.11575063 from
NNSFC (China), by grant No. 2015616 from the joint program in physics
between NSF and Binational (US-Israel) Science Foundation, and by grant No.
1286/17 from the Israel Science Foundation, Z.C. acknowledges an excellence
scholarship provided by the Tel Aviv University. B.A.M. appreciates
hospitality of the Joint Quantum Centre (JQC) Durham-Newcastle, during his
Visiting Professorship at Newcastle University, and NPP funding from EPSRC,
UK (Grant No. EP/K03250X/1).

%\bibliography{binaryTOF}
%\bibliographystyle{iop}

\providecommand{\newblock}{}

\end{document}